\documentclass[12pt]{article}
\usepackage{graphicx}
\usepackage{amsfonts}%
\usepackage{amssymb}%
\usepackage{mathrsfs}
\usepackage{amsmath}
\begin{document}
\renewcommand{\thefootnote}{\fnsymbol{footnote}}
\begin{titlepage}

\vspace{10mm}
\begin{center}
{\Large\bf Interpretation of the Cosmological Constant Problem within the Framework of Generalized Uncertainty Principle}
\vspace{16mm}

{\large Yan-Gang Miao${}^{1,2,}$\footnote{E-mail address: miaoyg@nankai.edu.cn} and Ying-Jie Zhao${}^{1,}$\footnote{E-mail address: xiangyabaozhang@mail.nankai.edu.cn}

\vspace{6mm}
${}^{1}${\normalsize \em School of Physics, Nankai University, Tianjin 300071, China}

\vspace{3mm}
${}^{2}${\normalsize \em Max-Planck-Institut f\"ur Physik (Werner-Heisenberg-Institut), F\"ohringer Ring 6,
\\80805 M\"unchen, Germany}}

\end{center}

\vspace{10mm}
\centerline{{\bf{Abstract}}}
\vspace{6mm}
We propose an improved exponential Generalized Uncertainty Principle (GUP) by introducing a positive integer $n$ called the suppressing index. Due to the UV/IR mixing brought by the GUP, the  states with momenta smaller than the critical momentum ($P < P_{\rm Crit}$) are canceled by the states with momenta larger than the critical momentum ($P > P_{\rm Crit}$) and thus have no contributions to the energy density of the vacuum. By considering the contributions just from the states with momenta larger than the critical momentum ($P > P_{\rm Crit}$) and choosing a suitable suppressing index, $n \sim 10^{123}$, we calculate the cosmological constant consistent with the experimentally observed value.

\vskip 20pt
\noindent
{\bf PACS Number(s)}: 98.80.Es, 02.40.Gh, 03.65.Sq


\end{titlepage}
\newpage
\renewcommand{\thefootnote}{\arabic{footnote}}
\setcounter{footnote}{0}
\setcounter{page}{2}

\section{Introduction}
The cosmological constant was introduced by Einstein~\cite{Einstein} in his field equations of general relativity in order to achieve a static universe, but abandoned after Hubble's discovery of the
expansion of the universe~\cite{Hubble}. Further observations of distance-redshift relations from the Type Ia supernovae~\cite{Riess} indicate that the expansion of the universe is accelerating, which requires a positive cosmological constant in order to produce a negative pressure. The most recent Planck 2013 results~\cite{Planck}  give an experimentally  observed small positive value of the cosmological constant, $\Lambda_{\rm Observed} \sim 10^{-47}\,  GeV^4/(\hbar^3 c^3)$. Meanwhile, the cosmological constant is interpreted as the energy density of the vacuum ($\rho_{vac}$) from the relation $\Lambda=\frac{8\pi G}{c^2}\rho_{vac}$, where $G$ is the gravitational constant and $c$ the speed of light,
and calculated to be $\Lambda_{\rm QFT}  \sim 10^{74}\, GeV^4/(\hbar^3 c^3)$ in quantum field theory (QFT),
which is about ${120}$ orders of the magnitude larger than the experimentally observed value. This is the well-known cosmological constant problem.

There have been some approaches to solve the problem, such as those based on the supersymmetry, supergravity, and superstrings, and on the anthropic consideration, the adjustment mechanisms, the changing gravity, and the quantum cosmology, etc., see, for instance, the review articles~\cite{WeinCarr}. In addition, a specific semiclassical approach~\cite{Zhang1}, based on the GUP with the quadratic momentum term~\cite{KMM}, makes an interesting attempt  although it does not solve the cosmological constant problem even in models with large extra dimensions~\cite{Extra}. In ref.~\cite{Zhang1}, the cosmological constant is rendered finite with the Planck mass ${M_{\rm Pl}} = \sqrt {{{\hbar c}}/{{{G}}}}\sim 10^{19}\,  GeV/c^2$ acting as a natural (not introduced by hand) UV cutoff brought by this type of GUP. Similarly, based on an exponential GUP~\cite{exp}, the cosmological constant problem cannot be solved, either, because the exponential GUP just provides an additional proportional factor of 2/9 to the result given by the quadratic GUP.

A GUP, as a phenomenological description of quantum gravity effects, can be constructed by deforming the Heisenberg uncertainty principle. It usually gives rise to a minimal length of the order of  the Planck length ${\ell_{\rm Pl}} = \sqrt {{{G\hbar }}/{{{c^3}}}}\sim 10^{-33}$ {\em cm} that is consistent with the fundamental length scale required by perturbative string theory~\cite{PST}.
The minimal length uncertainty relation then presents a fascinating property --- the UV/IR mixing:  A large $\Delta P$ (UV) corresponds to a large $\Delta X$ (IR). The UV/IR mixing as the characteristic of a minimal length uncertainty relation seems to be counterintuitive from the point of view of the Heisenberg uncertainty relation. Nonetheless, it is intimately  related to a wide range of physics, such as the AdS/CFT correspondence~\cite{AdSCFT}, the noncommutative field theory~\cite{NCFT}, the quantum gravity in asymptotically de Sitter spaces~\cite{deSitter}, and the inflationary cosmology~\cite{infcos}, etc.

The motivation of the present paper is to compute the cosmological constant and to ensure the consistency of our theoretical calculation and the experimentally observed value within the framework of GUP.
Through analyzing the problems existed in refs.~\cite{Zhang1,exp}, we make improvements in two aspects. On the one hand, we propose a new GUP because neither the quadratic nor the exponential GUP can provide enough suppression to the  energy density of the vacuum. On the other hand, because the UV/IR mixing results in~\cite{Zhang2} the cancellation of the states with momenta smaller  than the critical momentum ($P < P_{\rm Crit}$)  by the states with momenta larger than the critical momentum ($P > P_{\rm Crit}$) we regard this critical value as the lower limit of momentum integral rather than  zero in calculation of the energy density of the vacuum.  Following the semiclassical method described in ref.~\cite{Zhang1} together with our two improvements, we can then give  the cosmological constant consistent with the experimentally observed value.

The arrangement of the paper exactly presents our two improvements mentioned above. In the next section we propose an improved exponential GUP by introducing a positive integer $n$ in one-dimensional space and then extend it to $D$-dimensional space. Next, we give the analog of the Liouville theorem for this type of GUP in $D$ dimensions, i.e., the volume of phase space that is proved to be invariant under time evolution in the classical limit. In section 3 we calculate the energy density of the vacuum by focusing only on the states with momenta larger than the critical momentum due to the effect of UV/IR mixing brought by our improved exponential GUP.  We fix the integer introduced in our exponential GUP, $n\sim 10^{123}$, by requiring the consistency of our theoretical calculation and the experimental data of the cosmological constant. The last section is devoted to a brief conclusion.

\section{Improved exponential GUP and phase space volume in $D$ dimensions}
Our improved exponential GUP takes the form in one-dimensional space,
\begin{equation}
[ {\hat{X},\hat{P}}] = i\hbar \,{e^{\beta^{n} {\hat{P}^{2 n}}}},\label{expocr}
\end{equation}
where $\hbar \sqrt{\beta}$ is proportional to the Planck length, and $n$ is a positive integer to be determined. In the following of this section we investigate some general properties of this GUP for any $n$. In the next section, we shall fix $n$ by requiring that our calculation of the cosmological constant equals the experimentally observed value. We shall also see that  $n$ has the function to suppress the contributions of high momentum states to the energy density of the vacuum and thus it is named as the suppressing index.
 
The uncertainty relation corresponding to our improved exponential GUP reads,
\begin{eqnarray}
\left(\Delta X\right) \left(\Delta P\right) &\ge& \frac{\hbar }{2}\langle {{e^{\beta^n {\hat{P}^{2 n}}}}} \rangle
 = \frac{\hbar }{2}\sum_{k=0}^{\infty} {\frac{\beta ^{nk}}{k!}\langle {{\hat{P}^{2nk}}} \rangle } \nonumber \\
& \ge & \frac{\hbar }{2}{\sum_{k=0}^{\infty} {\frac{\beta ^{nk}}{k!}\langle {\hat{P}^2}} \rangle^{nk}}
 = \frac{\hbar }{2}{e^{\beta^n \langle {{\hat{P}^{2}}}\rangle^n }} \nonumber \\
& = &\frac{\hbar }{2}{e^{{\beta ^n}{{\left( {{\langle \hat P\rangle^2} + \left(\Delta P\right)^2} \right)}^n}}},\label{uncertainty}
\end{eqnarray}
where the properties $\langle \hat{P}^{2n} \rangle \ge \langle \hat{P}^{2} \rangle^n$ and $\langle {{\hat{P}^{2}}}\rangle={{\langle \hat P\rangle^2} + \left(\Delta P\right)^2}$ have been used.
Under the saturate condition $\langle \hat P\rangle=0$, we can get the minimal length $\left(\Delta X\right)_{\rm Min}=\frac{{\hbar \sqrt{\beta}}}{2}{\left( {2ne} \right)^{1/(2n)}}$ when $\Delta P$ takes the critical value,
$\Delta P={\left( \frac{1}{2n} \right)}^{1/(2n)}\frac{1}{{\sqrt{\beta}}}\equiv P_{\rm Crit}$. 
We can see that when $\left(\Delta X\right)_{\rm Min}$ is in the order of the Planck length, $\left(\Delta X\right)_{\rm Min} \sim \ell_{\rm Pl}$, $ P_{\rm Crit}$ must be in the order of the Planck momentum, $P_{\rm Crit} \sim P_{\rm Pl}=M_{\rm Pl}\,c$. Moreover, we note that the minimal length is always around the Planck length and the critical momentum is always around the Planck momentum for any $n$, which is a good property of our improved exponential GUP because $\left(\Delta X\right)_{\rm Min}$ can never tend to a macroscopic order of magnitude even for a quite great $n$, like  $n \sim 10^{123}$, see the next section. 

A natural generalization of eq.~(\ref{expocr}) in $D$ dimensions is
\begin{equation}
[ {{\hat X_i},{\hat P_j}} ] = i\hbar\,{e^{\beta^{n} {\hat P^{2n}}}}{\delta _{ij}},\label{gupext}
\end{equation}
where $i, j=1, 2, \cdots, D$, ${\hat P}^2 =\sum\limits_{i = 1}^D {{\hat P_i}^2}$,
and the components of the momentum operator $ \hat {\textbf{P}}$ are supposed to be commutative with each other,
\begin{equation}
[ {{\hat P_i},{\hat P_j}} ] = 0.\label{gupextpp}
\end{equation}
The commutation relations among the components of position operators are thus determined by the Jacobi identity as follows,
\begin{equation}
[ {{\hat X_i},{\hat X_j}} ] = 2i\hbar n{\beta ^n}{{\hat P}^{2\left( {n - 1} \right)}}{e^{{\beta ^n}{{\hat P}^{2n}}}}( {{\hat P_i}{\hat X_j} - {\hat P_j}{\hat X_i}} ).\label{gupextxx}
\end{equation}

Similar to the discussion in one-dimensional space, when the minimal length is in the order of the Planck length, the critical momentum must be in the order of the Planck momentum in $D$-dimensional space.

The analog of the Liouville theorem in the classical limit for our exponential GUP in $D$ dimensions gives the following
weighted phase space volume,
\begin{equation}
 {e^{ - D\beta^n {{P}^{2n}}}} {\mathrm{d}^D} \textbf{X} {\mathrm{d}^D}\textbf{P},
 \end{equation}
which is invariant under time evolution. For the proof, see Appendix.
Therefore, the density of states in momentum space  has the form,
\begin{equation}
\frac{1}{{{(2\pi\hbar)^D}}}{e^{ - D\beta^n {{P}^{2n}}}}{\mathrm{d}^D}\textbf{P}, \label{statedensity} 
\end{equation}
where the coordinates have been integrated over.  Note that the density of states in momentum space is modified by the damping factor $e^{ - D\beta^n {{P}^{2n}}}$, where the newly introduced suppressing index $n$ will take effect on suppression of the contributions of high momentum states to the energy density of the vacuum as expected. 

\section{The cosmological constant and suppressing index}
The energy density of the vacuum is
the sum of the zero-point energy  of each oscillator over momentum states per unit space volume. For each photon with vanishing rest mass, its zero-point energy equals $\frac{1}{2}\hbar \omega =\frac{1}{2}\sqrt{P^2c^2+m^2c^4}=\frac{1}{2}Pc$. Using eq.~(\ref{statedensity}) one gives the energy density of the vacuum in 3-dimensional space,
\begin{eqnarray}
\Lambda  &=&\frac{1}{(2\pi{\hbar})^3} \int {{\mathrm{d}^3}\textbf{P}}\, {e^{ - 3\beta^n {P^{2n}}}}\,\frac{1}{2}Pc \nonumber \\
 & =& \frac{c}{4 \pi ^2 {\hbar}^3}\int  {\mathrm{d}P}\, {P^3}\,{e^{ - 3\beta^n {P^{2n}}}}.\label{VED}
\end{eqnarray}
At present, the key point is to determine
the lower and upper limits of integral.  Under the usual Heisenberg uncertainty principle, all the states with momenta from zero to infinity are considered to contribute  to the energy density of the vacuum.

However, as stated in ref.~\cite{Zhang2}, not all momentum states have contributions to the energy density of the vacuum when the UV/IR mixing occurs. All the momentum states are divided into two parts, one part of states with momenta smaller than $P_{\rm Crit}$, and the other part of states with momenta large than $P_{\rm Crit}$.
The critical momentum $P_{\rm Crit}$ is just the momentum measurement precision that corresponds to the minimal length.
The first part called sub-Planckian modes  is canceled by the second part called trans-Planckian modes~\cite{Zhang2}, and thus it is prevented from contributing the energy density of the vacuum. Alternatively, this phenomenon can be explained by introducing a dynamical connection between the two parts of momentum states~\cite{Bank}.

For our exponential GUP, $P_{\rm Crit}\approx {\left( \frac{1}{2n} \right)}^{1/(2n)}\frac{1}{{\sqrt{\beta}}}$ in $D=3$ dimensions. Therefore, considering the contributions just from the trans-Planckian modes we can now determine that the lower limit of momentum integration is  this value instead of zero, while the upper limit is unchanged.  Eq.~(\ref{VED}) now reads
\begin{eqnarray}
\Lambda \approx \frac{c}{4 \pi ^2 {\hbar}^3} \int_{{\left( \frac{1}{2n} \right)}^{1/(2n)}\frac{1}{{\sqrt{\beta}}}}^{\infty} {\mathrm{d}P}\, {P^3}\,{e^{ - 3\beta^n {P^{2n}}}}.
\end{eqnarray}
Introducing a dimensionless parameter $x \equiv \beta {P^{2}}$, we rewrite the energy density of the vacuum and perform the integration as follows,
\begin{eqnarray}
\Lambda &\approx& \frac{ {\hbar}c}{8\pi ^2}\frac{1}{\left({\hbar \sqrt{\beta}}\right)^4} \int_{{\left( \frac{1}{2n} \right)}^{1/n}}^{\infty} {\mathrm{d}x}\, x\,{e^{ - 3x^n}},\nonumber \\
& =& \frac{ {\hbar}c}{8\pi ^2}\frac{1}{\left({\hbar \sqrt{\beta}}\right)^4}\frac{1}{3^{2/n}n}\,\Gamma\left(\frac{2}{n}, \frac{3}{2n}\right),
\end{eqnarray}
where the upper incomplete gamma function is defined as $\Gamma(s, z)= \int_{z}^{\infty}\mathrm{d}t\,t^{s-1}\,e^{-t}$.
Considering $\left(\Delta X\right)_{\rm Min} \sim {\hbar \sqrt{\beta}} \sim \ell_{\rm Pl}=\hbar/\left(M_{\rm Pl}\,c\right)$, we estimate the coefficient in front of the above definite integration, $\frac{ {\hbar}c}{8\pi ^2}\frac{1}{({\hbar \sqrt{\beta}})^4}\sim \frac{1}{8\pi ^2}\frac{(M_{\rm Pl}c^2)^4}{{\hbar}^3c^3}\sim 10^{74}\,GeV^4/(\hbar^3 c^3)$. Now requiring that our calculation is consistent with the experimentally observed value, i.e., $\Lambda=\Lambda_{\rm Observed}\sim 10^{-47}\,  GeV^4/(\hbar^3 c^3)$, we obtain
\begin{equation}
\frac{1}{3^{2/n}n}\,\Gamma\left(\frac{2}{n}, \frac{3}{2n}\right)\sim 10^{-121},
\end{equation}
which will give a very large suppressing index. The asymptotic behavior of the upper incomplete gamma function gives  $\lim_{s \to 0^{+}} \Gamma(s, z)= E_1(z)$, where $E_1(z)$ is the Expintegral function that is defined by $E_1(z)=\int_{1}^{\infty}\mathrm{d}t\,t^{-1}\,e^{-zt}$. Using the series of the Expintegral function, $E_1(z)=-\gamma-\ln z-\sum_{k=1}^{\infty}\frac{(-1)^k z^k}{k!\,k}$, where $\gamma$ is the Euler's constant, we estimate the asymptotic behavior of $zE_1(z)$: $ zE_1(z) \sim z$ when $z$ tends to a very small positive value. Thus we deduce $\frac{1}{3^{2/n}n}\,\Gamma\left(\frac{2}{n}, \frac{3}{2n}\right)\sim \frac{1}{n}$ when $n$ is very large. As a result, using the analysis of the asymptotic behaviors of the two special functions we estimate $n \sim 10^{121}$. A more precise estimation finally gives
\begin{eqnarray}
n \sim 10^{123}.\label{valueofn}
\end{eqnarray}
Consequently, we make an interpretation of the cosmological constant problem based on the improved exponential GUP (eqs.~(\ref{gupext})-(\ref{gupextxx})) with the above very large suppressing index.

\section{Conclusion}
We propose a simple and intuitive approach to interpret the cosmological constant problem within the framework of GUP --- a kind of phenomenological descriptions of quantum gravity. To this end, a new GUP is needed and its related UV/IR mixing effect should particularly be considered. The combination of the two aspects reaches our goal, that is, we reconcile the huge difference of about ${120}$ orders of magnitude ($\Lambda_{\rm QFT}/\Lambda_{\rm Observed}  \sim 10^{120}$) emerged between the QFT theoretical calculation and the experimentally observed value of the cosmological constant.

Finally, we make two comments. For the first, this extremely large suppressing index (eq.~(\ref{valueofn})) depends on both the ratio $\Lambda_{\rm QFT}/\Lambda_{\rm Observed}$Ó and our exponential GUP, and approximately equals the ratio. The quite large ratio gives rise to the quite large index that describes physically the strength of the UV/IR mixing effect. This characteristic can be seen clearly from eq.~(\ref{uncertainty}), that is, the larger $n$ is, the more rapidly $\Delta X$ grows when the momentum measurement precision continues to increase beyond the order of the Planck momentum. In other words, the larger $n$ is, the more efficiently the energy density of the vacuum (the cosmological constant) is suppressed through the UV/IR mixing effect. For the second comment, even though our calculation is consistent with the experimental data, our approach would be regarded as an interpretation that the origin of the cosmological constant problem may arise from the GUP issue rather than as a solution of the cosmological constant problem. 


\section*{Acknowledgments}
Y-GM would like to thank
D. L\"ust of the Max-Planck-Institut f\"ur Physik for kind hospitality where part of the work was performed.
This work was supported in part by the National Natural Science Foundation of China under grant No.11175090 and by the Ministry of Education of China under grant No.20120031110027. At last, the authors would like to thank the anonymous referee  for his/her helpful comments that improve this work greatly.

\section*{Appendix \hspace{.24cm}Invariance of the phase space volume  under time evolution}

\setcounter{equation}{0}
\renewcommand\theequation{A\arabic{equation}}

In the classical limit the commutation relations of operators eqs.~(\ref{gupext})-(\ref{gupextxx}) reduce to the Poisson brackets via $\frac{1}{i\hbar}[\hat{A}, \hat{B}] \to \{A, B\}$,
\begin{eqnarray}
\left\{ {{ X_i},{P_j}} \right\} &=& e^{\beta^{n} { P^{2n}}}{\delta _{ij}},\\
\left\{ {P_i},{ P_j} \right\} &=& 0, \\
\left\{ {{ X_i},{ X_j}} \right\} &=& 2 n{\beta ^n}{{P}^{2\left( {n - 1} \right)}}{e^{{\beta ^n}{{ P}^{2n}}}}( {{ P_i}{ X_j} - { P_j}{ X_i}} ),
\end{eqnarray}
where $i, j=1, 2, \cdots, D$, ${P}^2 =\sum\limits_{i = 1}^D {{P_i}^2}$.
Using the generic Poisson bracket of arbitrary functions,
\begin{equation}
\left\{ {F,G} \right\} = \left( {\frac{{\partial F}}{{\partial {X_i}}}\frac{{\partial G}}{{\partial {P_j}}} - \frac{{\partial F}}{{\partial {P_i}}}\frac{{\partial G}}{{\partial {X_j}}}} \right)\left\{ {{X_i},{P_j}} \right\} + \frac{{\partial F}}{{\partial {X_i}}}\frac{{\partial G}}{{\partial {X_j}}}\left\{ {{X_i},{X_j}} \right\},
\end{equation}
where  the Einstein summation convention has been used, we get the equations of motion of  $X_i$ and $P_i$,
\begin{eqnarray}
{{\dot X}_i} &=& \left\{ {{X_i},H} \right\} = \left\{ {{X_i},{P_j}} \right\}\frac{{\partial H}}{{\partial {P_j}}} + \left\{ {{X_i},{X_j}} \right\}\frac{{\partial H}}{{\partial {X_j}}}, \\
{{\dot P}_i} &=& \left\{ {{P_i},H} \right\} =  - \left\{ {{X_j},{P_i}} \right\}\frac{{\partial H}}{{\partial {X_j}}}.
\end{eqnarray}

During an infinitesimal time interval $\delta t$, the evolutions of $X_i$ and $P_i$ take the forms,
\begin{eqnarray}
{{X}_i}^\prime &=& {X_i} + {{\dot X}_i}\,\delta t = {X_i} + \left( {\left\{ {{X_i},{P_j}} \right\}\frac{{\partial H}}{{\partial {P_j}}} + \left\{ {{X_i},{X_j}} \right\}\frac{{\partial H}}{{\partial {X_j}}}} \right)\delta t, \\
{P_i}^\prime  &=& {P_i} + {{\dot P}_i}\,\delta t = {P_i} - \left\{ {{X_j},{P_i}} \right\}\frac{{\partial H}}{{\partial {X_j}}}\delta t,
\end{eqnarray}
and  the infinitesimal phase space volume changes  to be
\begin{equation}
{\mathrm{d}^D}\textbf{X}'{\mathrm{d}^D}\textbf{P}' = \left| {\frac{{\partial \left( {{X_1}^\prime, \cdots ,{{X}_D}^\prime;{{P}_1}^\prime, \cdots ,{P_D}^\prime} \right)}}{{\partial \left( {{X_1}, \cdots ,{X_D};{P_1}, \cdots ,{P_D}} \right)}}} \right|{\mathrm{d}^D}\textbf{X}{\mathrm{d}^D}\textbf{P},
\end{equation}
where the elements of the Jacobian are as follows,
\begin{eqnarray}
& & \frac{{\partial {{X}_i}^\prime}}{{\partial {X_j}}} = {\delta _{ij}} + \frac{{\partial \delta {X_i}}}{{\partial {X_j}}},\qquad
\frac{{\partial {{X}_i}^\prime}}{{\partial {P_j}}} = \frac{{\partial \delta {X_i}}}{{\partial {P_j}}}, \\
& & \frac{{\partial {{P}_i}^\prime}}{{\partial {X_j}}} = \frac{{\partial \delta {P_i}}}{{\partial {X_j}}},\qquad
\frac{{\partial {{P}_i}^\prime}}{{\partial {P_j}}} = {\delta _{ij}} + \frac{{\partial \delta {P_i}}}{{\partial {P_j}}}.
\end{eqnarray}

To the first order in $\delta t$, this Jacobian equals
\begin{equation}
\left| {\frac{{\partial \left( {{X_1}^\prime, \cdots ,{{X}_D}^\prime,{{P}_1}^\prime, \cdots ,{P_D}^\prime} \right)}}{{\partial \left( {{X_1}, \cdots ,{X_D},{P_1}, \cdots ,{P_D}} \right)}}} \right| = 1 + \frac{{\partial \delta {X_i}}}{{\partial {X_i}}} + \frac{{\partial \delta {P_i}}}{{\partial {P_i}}},
\end{equation}
and then we have
\begin{eqnarray}
\left( {\frac{{\partial \delta {X_i}}}{{\partial {X_i}}} + \frac{{\partial \delta {P_i}}}{{\partial {P_i}}}} \right)\frac{1}{{\delta t}} &= & \frac{\partial }{{\partial {X_i}}}\left( {\left\{ {{X_i},{P_j}} \right\}\frac{{\partial H}}{{\partial {P_j}}} + \left\{ {{X_i},{X_j}} \right\}\frac{{\partial H}}{{\partial {X_j}}}} \right) - \frac{\partial }{{\partial {P_i}}}\left( {\left\{ {{X_j},{P_i}} \right\}\frac{{\partial H}}{{\partial {X_j}}}} \right)    \nonumber \\
&=& \left( {\frac{\partial }{{\partial {X_i}}}\left\{ {{X_i},{X_j}} \right\}} \right)\frac{{\partial H}}{{\partial {X_j}}} - \left( {\frac{\partial }{{\partial {P_i}}}\left\{ {{X_j},{P_i}} \right\}} \right)\frac{{\partial H}}{{\partial {X_j}}}  \nonumber \\
&=&  - 2Dn{\beta ^n}{P^{2\left( {n - 1} \right)}}{e^{{\beta ^n}{P^{2n}}}}{P_i}\frac{{\partial H}}{{\partial {X_i}}}.
\end{eqnarray}
Moreover, to the first order in $\delta t$ the weight factor changes to be
\begin{eqnarray}
{e^{ - D\beta^n {{P'}^{2n}}}} &=& {e^{ - D\beta^n {{\left( {{P_i} + \delta {P_i}} \right)}^{2n}}}} \nonumber\\
&=& {e^{ - D\beta^n \left( {{P^{2}} - {P_i}\left\{ {{X_j},{P_i}} \right\}\frac{{\partial H}}{{\partial {X_j}}}\delta t - \left\{ {{X_j},{P_i}} \right\}\frac{{\partial H}}{{\partial {X_j}}}{P_i}\delta t} \right)^n}} \nonumber\\
&= &{e^{ - D\beta^n {P^{2n}}}}\left( {1 +  2Dn{\beta ^n}{P^{2\left( {n - 1} \right)}}{e^{{\beta ^n}{P^{2n}}}}{P_i}\frac{{\partial H}}{{\partial {X_i}}}\delta t} \right).
\end{eqnarray}
As a consequence,  we arrive at the result that the weighted phase space volume is invariant under time evolution,  i.e.,
\begin{eqnarray}
& &{e^{ - D\beta^n {{P'}^{2n}}}} {\mathrm{d}^D}\textbf{X}'{\mathrm{d}^D}\textbf{P}'  \nonumber \\
& =& {e^{ - D\beta^n {{P}^{2n}}}}\left( {1 +  2Dn{\beta ^n}{P^{2\left( {n - 1} \right)}}{e^{{\beta ^n}{P^{2n}}}}{P_i}\frac{{\partial H}}{{\partial {X_i}}}\delta t} \right)\nonumber \\
& & \times \left({1 -  2Dn{\beta ^n}{P^{2\left( {n - 1} \right)}}{e^{{\beta ^n}{P^{2n}}}}{P_j}\frac{{\partial H}}{{\partial {X_j}}}\delta t} \right){\mathrm{d}^D}\textbf{X} {\mathrm{d}^D}\textbf{P} \nonumber \\
& =& {e^{ - D\beta^n {{P}^{2n}}}} {\mathrm{d}^D} \textbf{X} {\mathrm{d}^D}\textbf{P}.
\end{eqnarray}

\end{document}